# Diversity-Multiplexing Tradeoff via Asymptotic Analysis of Large MIMO Systems

Sergey Loyka, George Levin
School of Information Technology and Engineering, University of Ottawa,
161 Louis Pasteur, Ottawa, Canada, K1N 6N5
E-mail: sergey.loyka@ieee.org

*Abstract*— Diversity–multiplexing tradeoff (DMT) presents a compact framework to compare various MIMO systems and channels in terms of the two main advantages they provide (i.e. high data rate and/or low error rate). This tradeoff was characterized asymptotically (SNR-> infinity) for i.i.d. Rayleigh fading channel by Zheng and Tse [1]. The asymptotic DMT overestimates the finite-SNR one [2]. In this paper, using the recent results on the asymptotic (in the number of antennas) outage capacity distribution, we derive and analyze the finite-SNR DMT for a broad class of channels (not necessarily Rayleigh fading). Based on this, we give the convergence conditions for the asymptotic DMT to be approached by the finite-SNR one. The multiplexing gain definition is shown to affect critically the convergence point: when the multiplexing gain is defined via the mean (ergodic) capacity, the convergence takes place at realistic SNR values. Furthermore, in this case the diversity gain can also be used to estimate the outage probability with reasonable accuracy. The multiplexing gain definition via the high-SNR asymptote of the mean capacity (as in [1]) results in very slow convergence for moderate to large systems (as $1/\ln(SNR)^2$) and, hence, the asymptotic DMT cannot be used at realistic SNR values. For this definition, the high-SNR threshold increases exponentially in the number of antennas and in the multiplexing gain. For correlated keyhole channel, the diversity gain is shown to decrease with correlation and power imbalance of the channel. While the SNR-asymptotic DMT of Zheng and Tse does not capture this effect, the size-asymptotic DMT does.

## I. INTRODUCTION

Multi-antenna (MIMO) systems are able to provide either high spectral efficiency (spatial multiplexing) or low error rate (high diversity) via exploiting multiple degrees of freedom available in the channel, but not both simultaneously as there is a fundamental tradeoff between the two. This tradeoff (DMT) is best characterized using the concepts of multiplexing and diversity gains [1]. Fundamentally, this is a tradeoff between the outage probability $P_{out}$, i.e. the probability that the fading channel is not able to support the transmission rate $R$, and the rate $R$, which can be expressed via the outage capacity distribution,

$$P_{out}(R) = \Pr[C < R] = F_C(R) \qquad (1)$$

where $C$ is the instantaneous channel capacity (i.e. capacity of a given channel realization), and $F_C(R)$ is its cumulative distribution function (CDF), also known as the outage capacity distribution. Defining the multiplexing gain $r$ as

$$r = \lim_{\gamma \to \infty} R / \ln\gamma \qquad (2)$$

where $\gamma$ is the average SNR at the receiver, and the diversity gain as[1]

$$d = -\lim_{\gamma \to \infty} \frac{\ln P_{out}}{\ln\gamma} \qquad (3)$$

the asymptotic ($\gamma \to \infty$) tradeoff for the independent identically distributed (i.i.d.) Rayleigh fading channel with the coherence time in symbols $l \geq m+n-1$ can be compactly expressed as [1],

$$d(r) = (n-r)(m-r), \; r = 0,1,...\min(m,n) \qquad (4)$$

where $m, n$ are the number of Tx, Rx antennas, for integer values of $r$, and using the linear interpolation in-between. The motivation for the definition of $r$ in (2) is that the mean (ergodic) capacity $\overline{C}$ scales as $\min(m,n)\ln\gamma$ at high SNR,

$$\overline{C} \approx \min(m,n)\ln\gamma, \text{ as } \gamma \to \infty \qquad (5)$$

and the motivation for the definition of $d$ in (3) is that $P_{out}$ scales as $\gamma^{-d}$ at high SNR,

$$P_{out} \approx c/\gamma^d, \text{ as } \gamma \to \infty \qquad (6)$$

where $c$ is a constant independent of the SNR.

While this approach provides a significant insight into MIMO channels and also into performance of various systems that exploit such channels, it has a number of limitations. Specifically, it does not say anything about operational significance of $r$ and $d$ at realistic (i.e. low to moderate) SNR. In other words, how high SNR is required to approach the asymptotes in (2),(3) with reasonable accuracy, so that, for example, $d$ can be used to accurately estimate $P_{out}$ using (6) and (4)? It was observed in [2], based on a lower bound to $P_{out}$ for Rayleigh and Rician channels, that the finite-SNR DMT lies well below the curve in (4), so that proper modifications to the asymptotic results and definitions are required for realistic SNR values. Using the asymptotic ($\gamma \to \infty$) DMT to compare two systems may give incorrect results at low to moderate SNR.

To evaluate the DMT for arbitrary SNR, one would need to known the outage capacity distribution $F_C(R)$. While some results of this kind are available in the literature, their complexity prevents any analytical development. A number of compact analytical results have recently appeared on the outage capacity distribution of asymptotically large systems, i.e. when either $n \to \infty$ or $m \to \infty$, or both [6]. For a broad class of channels (under mild technical conditions), it turns out to be Gaussian with the mean and the variance determined by specifics of the channel [3]-[6].

---
[1] while the original definition in [1] employed the average error rate, since it is dominated by the outage probability, the definition in (3) is equivalent to it. This definition has also been adopted in [2].

In this paper, we exploit these asymptotic results to derive the diversity-multiplexing tradeoff for arbitrary SNR and also for arbitrary-fading (i.e. not necessarily Rayleigh) i.i.d. channels. The advantage of this approach is that its results apply at any SNR and, thus, have operational significance at realistic SNR values. Our approach demonstrates that for moderately-large systems the convergence to the asymptotic (in SNR) results in [1] is very slow (as $1/(\ln\gamma)^2$). Furthermore, the asymptotic diversity gain in (3) alone cannot be used to estimate $P_{out}$ in (6) at any SNR (even very large) since the constant $c$ ("SNR offset") can be very large (e.g. $10^4$) for moderate to large systems. Thus, proper modifications of (2) and (3) are required to speed up the convergence in SNR, which are also presented in the paper.

Since it was demonstrated that the actual capacity distribution approaches the asymptotic (in system size) one already for a moderate number of antennas [3]-[6], our results also apply to the systems of realistic size.

The rest of the paper is organized as follows. In section II we introduce the basic system model, various assumptions and briefly review the asymptotic outage capacity distributions (Theorem 1 and 2), which is further used in section III to derive the finite-SNR DMT for arbitrary-fading i.i.d. and non-independent (correlated keyhole) channels. We also demonstrate, via Monte-Carlo simulations, that our asymptotic (in system size) results apply to moderate-size systems as well. The main results are summarized in Theorem 3, Corollaries 3.1 and 3.2 and eq. (33).

## II. SYSTEM MODEL AND OUTAGE CAPACITY DISTRIBUTION

The standard baseband discrete-time system model is adopted here,

$$\mathbf{r} = \mathbf{H}\mathbf{s} + \boldsymbol{\xi} \tag{7}$$

where $\mathbf{s}$ and $\mathbf{r}$ are the Tx and Rx vectors correspondingly, $\mathbf{H}$ is the $n\times m$ channel matrix, i.e. the matrix of the complex channel gains between each Tx and each Rx antenna, and $\boldsymbol{\xi}$ is the additive white Gaussian noise (AWGN), which is assumed to be $\mathcal{CN}(\mathbf{0}, \sigma_0^2 \mathbf{I})$, i.e. independent and identically distributed (i.i.d.) in each branch. The assumptions on the distribution of $\mathbf{H}$ follow those of the asymptotic capacity distributions (discussed in the next section): the entries of $\mathbf{H}$ are assumed to be either (i) i.i.d. but otherwise arbitrary fading (this includes Rayleigh fading as a special case) [6], which can also be extended to correlated identically distributed and independent non-identically distributed (the last two are not discussed in this paper due to the page limit) entries [9], or (ii) follow the statistics of the correlated keyhole channel [5].

When full channel state information (CSI) is available at the Rx end but no CSI at the Tx end, the instantaneous channel capacity (i.e. the capacity of a given channel realization $\mathbf{H}$) in nats/s/Hz is given by the celebrated log-det formula [7],

$$C = \ln\det\left(\mathbf{I} + \frac{\gamma}{m}\mathbf{H}\mathbf{H}^+\right) \tag{8}$$

where $\gamma$ is the average SNR per Rx antenna (contributed by all Tx antennas), "$+$" denotes conjugate transpose.

For large $m,n$, the distribution of $C$ takes on a remarkably simple form in a number of cases[2]:

**Theorem 1** [[6], Theorem 2.76]: Let $\mathbf{H}$ be an $n\times m$ channel matrix whose entries are i.i.d. zero mean random variables with unit variance such that $E[|H_{ij}|^4] = 2$. As both $m,n \to \infty$ and $\beta = m/n$ is a constant, the instantaneous capacity in (8) is asymptotically (in $m,n$) Gaussian, with the following mean $\overline{C}$ and variance $\sigma_C^2$:

$$\frac{\overline{C}}{n} = \beta\ln\left(1 + \frac{\gamma}{\beta} - \frac{1}{4}F\left(\frac{\gamma}{\beta},\beta\right)\right) + \ln\left(1 + \gamma - \frac{1}{4}F(\frac{\gamma}{\beta},\beta)\right) - \frac{\beta}{4\gamma_0}F\left(\frac{\gamma}{\beta},\beta\right) \tag{9}$$

$$\sigma_C^2 = -\ln\left(1 - \beta\left[\frac{1}{4\gamma}F\left(\frac{\gamma}{\beta},\beta\right)\right]^2\right) \tag{10}$$

where $F(x,z) = (\sqrt{x(1+\sqrt{z})^2 + 1} - \sqrt{x(1-\sqrt{z})^2 + 1})^2$.

**Theorem 2** [[5], Theorem 1]: Let $\mathbf{H} = \mathbf{h}_r\mathbf{h}_t^+$ be an $n\times m$ keyhole channel matrix, where $\mathbf{h}_t$ [$m\times 1$] and $\mathbf{h}_r$ [$n\times 1$] are mutually independent complex circular symmetric Gaussian random vectors representing the gains from the transmit antennas to the keyhole and from the keyhole to the receive antennas respectively. As both $m,n \to \infty$, the distribution of $C$ is asymptotically Gaussian if $\lim_{m\to\infty} m^{-1}tr(\mathbf{R}_t)$ and $\lim_{n\to\infty} n^{-1}tr(\mathbf{R}_r)$ are finite and $\lim_{m\to\infty} m^{-2}\|\mathbf{R}_t\|^2 = 0$, $\lim_{n\to\infty} n^{-2}\|\mathbf{R}_r\|^2 = 0$, where $\mathbf{R}_t = E(\mathbf{h}_t\mathbf{h}_t^+)$, $\mathbf{R}_r = E(\mathbf{h}_r\mathbf{h}_r^+)$ are the Tx and Rx end correlation matrices, and $\|\ \|$ denotes the Frobenius norm. If the channel is normalized so that $m^{-1}tr(\mathbf{R}_t) = 1$, $n^{-1}tr\{\mathbf{R}_r\} = 1$, the mean and the variance are asymptotically as follows:

$$\overline{C} = \ln(1 + n\gamma) \tag{11}$$

$$\sigma_C^2 = m^{-2}\|\mathbf{R}_t\|^2 + n^{-2}\|\mathbf{R}_r\|^2 \tag{12}$$

Using the asymptotic distributions above, the outage probability can be expressed as

$$P_{out}(R) = Q\left(\frac{\overline{C}-R}{\sigma_C}\right) \leq \frac{1}{2}\exp\left(-\frac{1}{2}\left(\frac{\overline{C}-R}{\sigma_C}\right)^2\right) \tag{13}$$

where $Q(x) = \frac{1}{\sqrt{2\pi}}\int_x^\infty \exp(-t^2/2)dt$. The upper bound in (13) becomes tight at moderate SNR, so we use it as an approximation to $P_{out}$ to simplify calculations.

## III. FINITE-SNR DMT VIA ASYMPTOTIC CAPACITY DISTRIBUTIONS

Finite-SNR DMT analysis requires using finite-SNR analogs of the definitions in (2),(3),

$$r = \frac{R}{\ln\gamma},\quad d_\gamma = -\frac{\ln P_{out}}{\ln\gamma} \tag{14}$$

The convergence of the finite-SNR DMT to the asymptotic one in (4) is significantly improved if $r$ is defined via $\overline{C}$, or via $\ln(\gamma/e)$, which is motivated by (18) and takes into account the high-SNR offset[3] $1/e$,

$$r = \frac{\min(m,n)R}{\overline{C}} \quad (15)$$

$$r = \frac{R}{\ln(\gamma/e)} \quad (16)$$

where (15) defines the rate as the $r/\min(m,n)$ fraction of the mean capacity.

---

[2] Other asymptotic results are also available in the literature. However, we will rely only on these two theorems in this paper.

Another possible definition of $d$, which was introduced in [2], captures the differential effect of diversity, i.e. how much increase in SNR is required to decrease $P_{out}$ by certain amount,

$$d'_\gamma = -\frac{\partial \ln P_{out}}{\partial \ln \gamma} \quad (17)$$

Note that the differential diversity gain $d'_\gamma$ is insensitive to the constant $c$ in (6) so that the convergence to the asymptotic value is faster. For high SNR, both definitions of the diversity gain (in (17) and (14)) give the same result.

While the diversity gain provides some indication of the performance, its usefulness lies in its relation with the outage probability (or the average error rate) as the latter is the ultimate performance indicator, not the diversity gain itself. Using the three multiplexing gain definitions in (14)-(16), Fig. 1 and 2 compare the outage probability vs. SNR from the asymptotic result in (13) to Monte-Carlo (MC) simulations for i.i.d. Rayleigh channel, which shows good agreement between the two (even for small system size, $n=2$). Note the anomalous behavior of the outage probability (increasing with the SNR) for the multiplexing gain definitions in (14), (16), which is due to the fact that the rate $R < \overline{C}$ on the corresponding interval but it increases faster than $\overline{C}$ with the SNR so that $|\overline{C}-R|/\sigma_C$ decreases; after the anomalous region this tendency is reversed. This never happens if the rate is defined as a fraction of the mean capacity (i.e. (15)).

Also note a high SNR offset ($c \approx 10^4$, see (6)) in $P_{out}$ for $R = r\ln\gamma$ and $n=10$. This makes it impossible to estimate $P_{out}$ from the diversity gain alone, i.e. using $P_{out} \approx 1/\gamma^d$, no matter how high the SNR is. The rough estimation $P_{out} \approx 1/\gamma^d$ works only if $c$ is on the order of unity. When this is not the case, $c$ has to be accounted for as well. This indicates the limitation of the DMT, which ignores the constant $c$. Specifically, when two systems (or channels) are compared with the same $r$, and $d_1 > d_2$, it does not mean that system 1 performs better than system 2 in terms of $P_{out}$ (or average error rate) since it may be that $c_1 > c_2$ and the latter effect is dominant. Hence, using the DMT curves alone to compare two systems may produce incorrect results, even at very high SNR. This suggests that the constant $c$ (high-SNR offset) should also be included in the DMT if the error rate performance is of importance. This problem is somewhat eliminated by using the multiplexing gain definition in (16), as $c$ becomes a moderate constant, but the anomalous behavior of the outage probability is not eliminated so that its estimation from the diversity gain alone at $\gamma \leq 30dB$ is not possible. Using the definition in (15) eliminates most of the problem, leaving only the moderate offset $c \approx 1/5$. For smaller systems (Fig. 2), this problem is not that severe (the SNR offset disappears at $\gamma \geq 15dB$), but the anomalous behavior of the outage probability at low to moderate SNR for all definitions of the multiplexing gain but in (15) is still present.

---
[3] [8] gives a detailed discussion of the importance of high-SNR offset in the capacity analysis of MIMO systems. Note that this offset is missing in (5).

We analyze below the finite SNR DMT analytically using the multiplexing gain definitions in (14)-(16) to clarify their advantages and disadvantages when applied to realistic systems (low to moderate SNR, moderate or small system size).

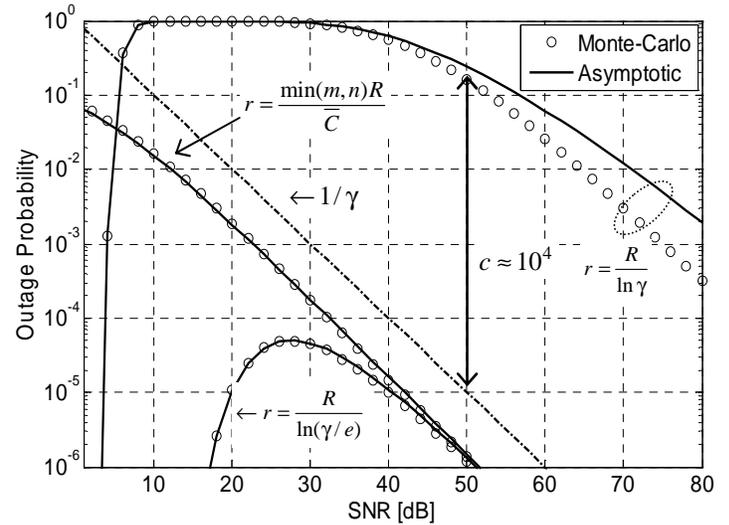

**Fig. 1. Outage probability vs. SNR for various definitions of the multiplexing gain; $n=m=10, r=9$; solid line – asymptotic from (9),(10),(13), circles – Monte-Carlo simulations ($10^9$ trials); dash line - $P_{out} = 1/\gamma$. Note high SNR offset ($c \approx 10^4$).**

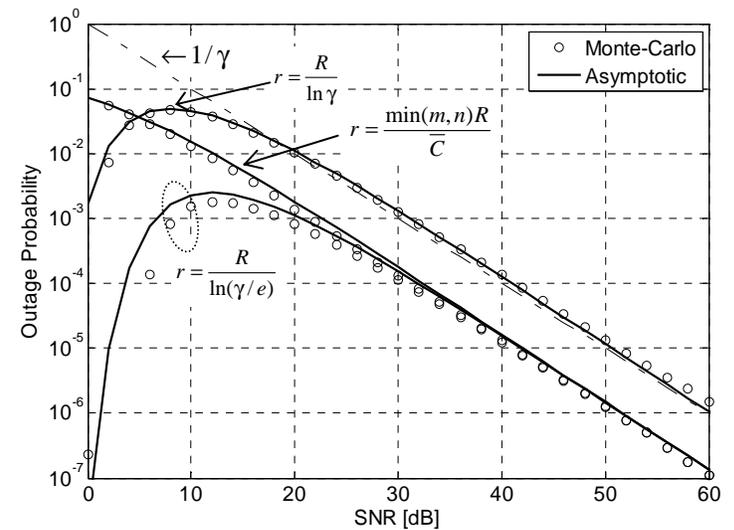

**Fig. 2. Outage probability vs. SNR for various definitions of the multiplexing gain; $n=m=2, r=1$; solid line – asymptotic from (9),(10),(13), circles – Monte-Carlo simulations ($10^8$ trials); dash line - $P_{out} = 1/\gamma$. The SNR offset is small in this case ($c \approx 1$) and the convergence is achieved at realistic SNR.**

## A. Independent Identically Distributed Channels

We begin with Theorem 1 and consider square channels, $\beta = 1$[4]. At moderate to high SNR, the mean and the variance can be approximated as[5]

$$\frac{\overline{C}}{n} = \ln\left(\frac{\gamma}{e}\right) + \frac{2}{\sqrt{\gamma}}, \quad \sigma_C^2 = \frac{1}{2}\left(\ln\frac{\gamma}{4} + \frac{2}{\sqrt{\gamma}}\right) \quad (18)$$

Numerical evaluation of (9), (10) indicates that (18) becomes accurate already for $\gamma \geq 5dB$.

To simplify the analysis and to get some insight, we use below high but finite SNR approximations, i.e. $\gamma \gg 1$ but not $\gamma \to \infty$. This approximations, as it is demonstrated below, hold true already at low or moderate SNR levels and allow one to quantify the effect of SNR on the DMT and, in particular, to establish the SNR levels at which the asymptotic results in [1] hold.

Substituting (18) into the upper bound in (13), using the multiplexing gain definition in (15), after some manipulations keeping only the lower-order (dominating) terms, one obtains

$$P_{out} \approx \frac{1}{2}\left(\frac{\gamma}{e}\right)^{-d(r)\Delta(\gamma)} \quad (19)$$

where $d(r) = (n-r)^2$ (as in (4)), and $\Delta(\gamma)$ quantifies the effect of finite SNR,

$$\Delta(\gamma) \approx 1 + 2/\left(\sqrt{\gamma}\ln(\gamma/e)\right) \quad (20)$$

Interpreting the $1/e$ term in (19) as a high-SNR offset (similarly to [8]), the diversity gain in (14) becomes $d_\gamma \approx d(r)\Delta(\gamma)$. Using (19), the differential diversity gain (17) can be expressed as

$$d'_\gamma = d(r)(\Delta(\gamma) + \gamma\ln(\gamma/e)\partial\Delta(\gamma)/\partial\gamma) \quad (21)$$

which, after some manipulations, can be simplified to

$$d'_\gamma \approx (n-r)^2\left(1 - \frac{1}{2\sqrt{\gamma}}\right) \quad (22)$$

The first factor in (22) is identical to (4) (recall that $m = n$), and the second term represents the effect of the finite SNR. As Fig. 3,4 demonstrate, (22) is reasonably accurate for $\gamma \geq 0dB$. The convergence to the asymptotic ($\gamma \to \infty$) result in (4) is achieved when the second term in (22) can be neglected, which we set, somewhat arbitrary, as $1/(2\sqrt{\gamma}) \leq 0.1$ (i.e. within 10% accuracy),

$$\gamma \geq 25 \approx 14dB \quad (23)$$

To indicate the impact of the rate definition on the convergence speed, the results above should be contrasted to those obtained using the other two definitions of the multiplexing gain in (14) and (16) respectively,

$$d'_\gamma \approx (n-r)^2\left(1 - \frac{n+r}{n-r}\frac{1}{\sqrt{\gamma}} - \left(\frac{r}{n-r}\right)^2\frac{1}{\ln(\gamma/e)^2}\right) \quad (24)$$

$$d'_\gamma \approx (n-r)^2\left(1 - \frac{n+r}{n-r}\frac{1}{\sqrt{\gamma}}\right) \quad (25)$$

These equations hold for $r < n$. If $r = n$, then $d_\gamma = 0$ and also $d'_\gamma = 0$, as it should be. Note that, as $\gamma \to \infty$, $d'_\gamma$ converges to the asymptote (4) for all multiplexing gain definitions. The convergence in (24) and (25) respectively is achieved for

$$\gamma \geq \max\left[\left(\frac{10(n+r)}{n-r}\right)^2, \exp\left(1 + \frac{3r}{n-r}\right)\right] \text{ (eq. 24)} \quad (26)$$

$$\gamma \geq \left(\frac{10(n+r)}{n-r}\right)^2 \text{ (eq. 25)} \quad (27)$$

Fig. 3 and 4 compare the differential diversity gain evaluated via the asymptotic distribution with the moments in (9), (10) to the approximations in (22), (24) and (25). Clearly, the approximations in (22), (24) and (25) are of reasonable accuracy.

The slowest convergence (i.e. logarithmic, as $1/(\ln\gamma)^2$) is for the multiplexing gain definition in (14), which was used in [1], and the fastest convergence is for the multiplexing gain definition in (15), which is also independent of any system parameters.

*Example 1*: convergence conditions for $n = 10$, $r = 9$,

$$\gamma \geq 50dB \text{ (the multiplexing gain in (16))} \quad (28)$$

$$\gamma \geq 120dB \text{ (the multiplexing gain in (14))} \quad (29)$$

Few observations are in order, based on (23),(28),(29): (i) the original multiplexing gain definition in (14) results in extremely slow convergence, making the results inapplicable at realistic SNR values; (ii) the high-SNR offset in (16) improves the convergence significantly, but yet not enough to achieve realistic SNRs; (iii) the multiplexing gain definition in (15) is the best, with the convergence at realistic SNR values.

To observe the effect of system parameters, consider another example.

*Example 2*: convergence conditions for $n = 2$, $r = 1$,

$$\gamma \geq 22dB \text{ (the multiplexing gain in (14) and (16))} \quad (30)$$

Comparing to Example 1, one concludes that the convergence for the multiplexing gains in (14) and (16) is significantly affected by the system size: for small systems, all three definitions give roughly the same (fast) convergence, achieved at realistic SNRs; for larger systems, only the definition in (15) results in convergence at realistic SNRs, which is also independent of the system size and rate. For the definition in (14) (which was used in [1]) the high-SNR threshold increases exponentially in system size and in the multiplexing gain (see (26)). Based on these observations, the multiplexing gain definition in (15) relying on the mean capacity seems to be the best one.

The main results of this section are summarized in the following Theorem and Corollaries.

---

[4] The results can also be generalized to arbitrary $\beta$, which is omitted here due to the page limit.

[5] similar approximations, without $2/\sqrt{\gamma}$ term, can be found elsewhere in the literature. They, however, become accurate for significantly larger SNR, $\gamma \geq 20...30dB$.

**Theorem 3**. Consider a fading channel satisfying the conditions of Theorem 1 with $n = m$. The finite-SNR diversity-multiplexing tradeoff using the diversity gain definition in (17) and the multiplexing gain definitions in (14), (15), (16), are given by (24), (22), (25), and the convergence to the asymptote ($\gamma \rightarrow \infty$) in (4) is achieved under the conditions in (26), (23), (27) respectively.

**Corollary 3.1:** Convergence of the finite-SNR DMT to the asymptotic ($\gamma \rightarrow \infty$) one is the fastest for $R = r\overline{C}/n$ and the slowest for $R = r \ln \gamma$. For moderate to large system size, only the former results in convergence at realistic SNR values.

**Corollary 3.2**: Only for $R = r\overline{C}/n$ the outage probability can be estimated from (6) using the diversity gain in (4), when $r$ is not too small. The other definitions in (14) and (16) result in large SNR offset and anomalous behavior of $P_{out}(\gamma)$ at realistic SNR values, for moderate to large system size.

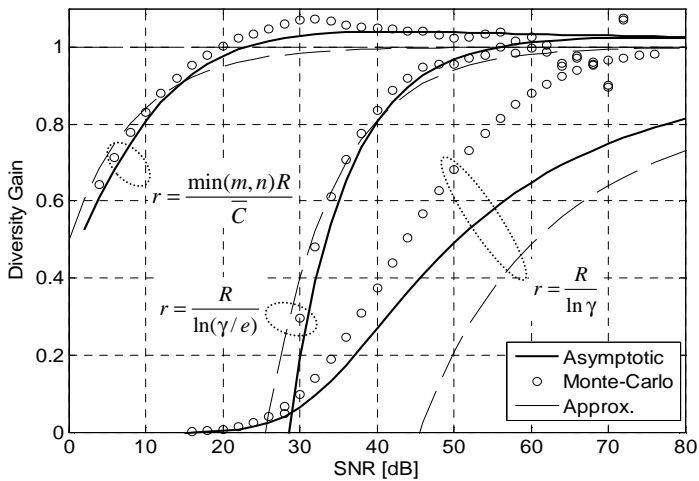

**Fig. 3. Differential diversity gain vs. SNR for various definitions of the multiplexing gain;** $n = m = 10, r = 9$; **solid line – asymptotic from (9),(10),(13), dashed – approximations in** (22), (24), (25).

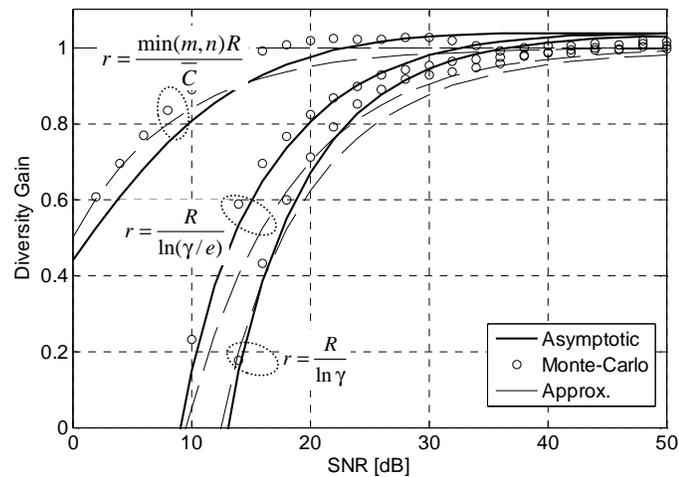

**Fig. 4. Differential diversity gain vs. SNR for various definitions of the multiplexing gain;** $n = m = 2, r = 1$.

*B. Correlated Keyhole Channel*

Using Theorem 2, similar results can also be obtained for correlated keyhole channels. Specifically, using the multiplexing gain definition in (15), the outage probability can be expressed as

$$P_{out} \approx \frac{1}{2}(n\gamma)^{-d(r)\Delta(\gamma)} \qquad (31)$$

where $r \leq 1$ and

$$d(r) = (1-r)^2, \; \Delta(\gamma) = \frac{\ln(\gamma n)}{2(m^{-2}\|\mathbf{R}_t\|^2 + n^{-2}\|\mathbf{R}_r\|^2)} \qquad (32)$$

Note a different SNR offset in (31) compared to (19). The differential diversity gain can be expressed as

$$d'_\gamma = \frac{(1-r)^2 \ln(\gamma n)}{m^{-2}\|\mathbf{R}_t\|^2 + n^{-2}\|\mathbf{R}_r\|^2} \qquad (33)$$

Eq. (33) demonstrates the effect of SNR and of the correlation on the finite-SNR DMT. The denominator in (33) is in fact the measure of correlation and power imbalance in a MIMO channel introduced in [10]. Thus, any correlation or power imbalance, at either Tx or Rx end, reduce the differential diversity gain.

Due to the asymptotic nature of the capacity distribution in Theorem 2, this result cannot be extended to $\gamma \rightarrow \infty$ for finite $n, m$ because of slow convergence (with $n, m$) of the distribution tail. However, it does provide a good approximation at moderate SNR values.